\documentclass[aps,pra,reprint,onecolumn,notitlepage,eqsecnum,floatfix,showkeys,nofootinbib]{revtex4-2}

\usepackage{amsmath}
\usepackage{amsbsy}
\usepackage{amsfonts}
\usepackage{xcolor}
\usepackage{graphicx}
\usepackage{hyperref}
\usepackage{multirow}
\usepackage{linearb}
\usepackage{relsize}
\usepackage{amsthm}
\usepackage{calrsfs}
\DeclareMathAlphabet\mathbfcal{OMS}{cmsy}{b}{n}

\newcommand{\bq}{\begin{eqnarray}}
\newcommand{\eq}{\end{eqnarray}}
\newcommand{\bqn}{\begin{eqnarray*}}
\newcommand{\eqn}{\end{eqnarray*}}
\newcommand{\bqs}{\begin{subequations}}
\newcommand{\eqs}{\end{subequations}}
\newcommand{\bw}{\begin{widetext}}
\newcommand{\ew}{\end{widetext}}

\newcommand{\rr}{{\boldsymbol r}}

\newcommand{\ddelta}{{\boldsymbol\Delta}}

\newcommand{\calo}{{\cal O}}

\newcommand{\calz}{{\cal Z}}

\newcommand{\calp}{{\cal P}}

\newcommand{\cale}{{\cal E}}
\newcommand{\calv}{{\cal V}}
\newcommand{\calk}{{\cal K}}

\begin{document}
\title{Quantum Hard Spheres with Affine Quantization}

\author{Riccardo Fantoni}
\email{riccardo.fantoni@scuola.istruzione.it}
\affiliation{Universit\`a di Trieste, Dipartimento di Fisica, strada
  Costiera 11, 34151 Grignano (Trieste), Italy}

\date{\today}

\begin{abstract}
We study a fluid of quantum hard-spheres treated with affine-quantization.
Assuming that the fluid obeys to Bose-Einstein statistics we solve for its
thermodynamic properties using the path integral Monte Carlo method.
\end{abstract}


\maketitle
\section{Introduction}
\label{sec:intro}

The simplest model of a fluid \cite{Hansen} is a system of hard spheres, for 
which the pair potential $\phi_{\rm HS}(r)$ at a separation $r$ is
\bq \label{eq:hsp}
\phi_{\rm HS}(r)=\left\{\begin{array}{ll}
+\infty & r<\sigma \\
0       & \mbox{else}
\end{array}
\right.,
\eq
where $\sigma$ is the hard-sphere diameter. This mathematical expression just 
describes the physical presence of an {\sl excluded volume} around the center 
of each hard-sphere particle. This simple potential is ideally 
suited to the study of phenomena in which the hard core of the potential is 
the dominant factor. Much of our understanding of the properties of the hard-sphere 
model come from computer simulations \cite{Alder1957,Alder1959}.
Such calculations have revealed very clearly that the structure of a hard-sphere 
fluid does not differ in any significant way from that corresponding to more 
complicated interatomic potentials, at least under conditions close to 
crystallisation.

The most important feature of the pair potential of a liquid is the strong repulsion 
that appears at short range and is due to the overlap of the outer electron shells 
inhibited by the Pauli exclusion principle. This strongly repulsive forces are 
responsible for the short range order characteristic of the liquid state. The 
attractive forces acting at long range are much more smooth and play only a minor 
role in determining the structure of the liquid. They provide an almost uniform
attractive background giving rise to the cohesive energy that stabilizes the
liquid.

However, although simulations show that the hard-sphere {\sl classical} fluid 
undergoes a freezing transition at a reduced number density 
$\rho^\star=\rho\sigma^3\approx 0.945$ \cite{Wood1957,Alder1957}, the absence 
of attractive forces means that there is only one fluid phase.
The properties of hard-spheres are better understood in terms of the so called
packing fraction $\eta=\rho v_d\sigma^d$ where
\bq
v_d=\frac{(\pi/4)^{d/2}}{\Gamma(1+d/2)},
\eq
is the volume of a $d$-dimensional sphere of unit diameter (for instance 
$v_3=\pi/6$, so that $0<\eta<v_3\sqrt{2}$ since one cannot pack spheres more
than their closest packing configuration in three dimensions).

The hard-sphere fluid is also considered as the reference system for a 
pair potential perturbation theory 
\cite{Andersen1971a,Andersen1971b,Andersen1972,Hansen}.

Least but not least the classical hard-sphere fluid admits exact analytic solution 
for the Percus-Yevick integral equation theory through the Wiener-Hopf 
factorization \cite{Thiele1963,Wertheim1963,Wertheim1964,Baxter1968,Baxter1970,Santosb}  

Now the {\sl quantum} version of the hard-sphere fluid has seldomly been treated.
This can be explained by the difficulties that one faces when treating hard-walls 
in quantum mechanics \cite{Fantoni23b}. In order to overcome the difficulty of 
having to deal with a canonical momentum that is not anymore a self-adjoint 
operator a possible way out strategy is to use {\sl affine quantization} instead of
the usual Dirac canonical quantization \cite{FantoniCQ}. We will then substitute
the usual ill defined canonical momentum operator with the self-adjoint 
{\sl dilation} operator in affine quantization, in order to describe correctly the 
excluded volume in the quantum realm.

In this work we will define the model of Quantum Hard-Spheres treated with 
Affine Quantization (AQHS) and then we will carry out some Path Integral Monte 
Carlo (PIMC) simulations for a canonical ensemble of the fluid of AQHS in 
thermal equilibrium at finite non-zero temperature. 

\section{The model}
\label{sec:model}

The Hamiltonian for a $d$-dimensional system of $N$ Bose (BHS) Quantum (QHS) 
Hard-Spheres (HS) treated with Affine-Quantization (AQ) is as follows 
\cite{Fantoni24b}
\bq \label{eq:H}
H=\lambda\left[-\sum_{i=1}^N\left(\frac{\partial}{\partial \rr_i}\right)^2
+\sum_{i<j}\phi_{\rm AQ}(r_{ij})\right]
+\sum_{i<j}\phi_{\rm HS}(r_{ij}),
\eq
where $\sigma$ is the diameter of the spheres of mass $m$, 
$\lambda=\hbar^2/2m$, $\rr_i$ is the position of the $i$th sphere center,
$\rr_{ij}=\rr_i-\rr_j$, the HS pair potential $\phi_{\rm HS}(r)$ of Eq.
(\ref{eq:hsp}), and the AQ effective pair potential 
\bq
\phi_{\rm AQ}(r)=\frac{2r^2+\sigma^2}{(r^2-\sigma^2)^2}.
\eq
The latter has a repulsive spike at contact $r=\sigma$. So that the QHS 
repel each other before getting into (classical) contact. And at large $r$,
$\phi(r)\approx a^2/r^2$, with $a=\hbar\sqrt 2$, which has a long range
repulsive character in $d>1$. So, in order for the system to admit a thermodynamic 
limit one needs to add a uniform background with the effect to modify the 
AQ pair potential as follows
\bq \label{eq:lrc-a}
\phi_{\rm AQ}(r)&\to&\phi_{\rm AQ}(r)-
\frac{N}{N-1}D_d,\\ \label{eq:lrc-b}
D_d&=&\frac{1}{\Omega}\int_{\Omega,r>\sigma}\phi_{\rm AQ}(r)\,d\rr
\approx\left\{\begin{array}{ll}
24(L-\sigma)/L^3    & d=3\\
16\ln(L/\sigma)/L^2 & d=2
\end{array}\right.
\eq
where $\Omega=L^d$ is the volume of the cubic box (approximated to a sphere 
of radius $L/2$ in Eq. (\ref{eq:lrc-b}) containing the system of 
QHS in thermal equilibrium at an inverse temperature $\beta=1/k_BT$ with 
$k_B$ the Boltzmann constant and $T$ the absolute temperature. This procedure 
is justified by requiring that the total potential energy
$\widehat{V}=\int d\rr d\rr'\,\rho(\rr)\rho(\rr')\phi_{\rm AQ}(|\rr-\rr'|)$
remains unchanged under the transformation (\ref{eq:lrc-a}) keeping 
the local number density $\rho(\rr)=\sum_i\delta^d(\rr-\rr_i)$, where 
$\delta^d$ is the $d$-dimensional Dirac delta function, or
under the transformation $\rho(\rr)\to\rho(\rr)+\rho_b$, where 
$\rho_b=-\rho=-N/\Omega$ is the uniform background density, keeping 
the pair potential $\phi_{\rm AQ}(\rr)$ constant.

\section{The Path Integral Monte Carlo method}
\label{sec:pimc}

The basis for our quantum simulation is imaginary time path integral 
\cite{Feynman1953}. We will treat the $N$ AQHS as bosons interacting with
the Hamiltonian of Eq. (\ref{eq:H}). Then the quantum statistical mechanics 
problem reduces to use path integral \cite{Feynman-Hibbs} to calculate the 
thermal {\sl density matrix} elements $\langle R|e^{-\beta H}|R'\rangle$ as 
the following path average
\bq
\langle R|e^{-\beta H}|R'\rangle=
\int dR_1\int dR_2\cdots\int dR_{M-1}
\,e^{-S(R_0,R_1,R_2,\ldots,R_M)},
\eq
where $R_i=(\rr_{1,i},\rr_{2,i},\ldots,\rr_{N,i})$ for $i=1,2,\ldots,M$ are the
``beads'' attached to the $M$ timeslices that form the ``polymer'' of the
Feynman ``classical-quantum'' isomorphism \cite{Ceperley1995}. 
And $R_0=R=(\rr_1,\rr_2,\ldots,\rr_N)$, $R_M=R'=(\rr'_1,\rr'_2,\ldots,\rr'_N)$.
In the limit that the {\sl timestep} $\tau=\beta/M\to 0$ the action $S$ has the 
following simple ``primitive approximation'' form \cite{Ceperley1995}
\bq
S(R_0,R_1,R_2,\ldots,R_M)=
\sum_{i=1}^M\left(\sum_{k=1}^N
\frac{(\rr_{k,i}-\rr_{k,i-1})^2}{4\lambda\tau}+
\tau\sum_{k<l}\left[\lambda\phi_{\rm AQ}(r_{kl,i})+\phi_{\rm HS}(r_{kl,i})\right]\right)
+\frac{dNM}{2}\ln(4\pi\lambda\tau),
\eq
where $r_{kl,i}=|\rr_{k,i}-\rr_{l,i}|$ and in $\rr_{i,j}$ the first index $i$
is a particle label and the second index $j$ is a bead label. We then measure 
an observable $\calo$ through the following trace
\bq
\langle\calo\rangle=\frac{\frac{1}{N!}\sum_{\calp}\int dR\,
\langle R|\calo e^{-\beta H}|\calp R\rangle}
{\frac{1}{N!}\sum_{\calp}\int dR\,
\langle R|e^{-\beta H}|\calp R\rangle}, 
\eq
where the sum is over all permutations of particle positions $\calp R$.
If $\calo$ is diagonal in position representation
\bq \label{eq:Omeasure}
\langle\calo\rangle=\frac{\frac{1}{N!}\sum_{\calp}\int dR\,
\calo(R)\langle R|e^{-\beta H}|\calp R\rangle}
{\frac{1}{N!}\sum_{\calp}\int dR\,
\langle R|e^{-\beta H}|\calp R\rangle}. 
\eq
We perform the $dNM$ multidimensional integral and the permutation sum with 
the {\sl Metropolis algorithm} \cite{Metropolis,Kalos-Whitlock}
\footnote{Note that a brute force integration with Monte Carlo is not 
feasible here because the integrand of Eq. (\ref{eq:Omeasure}) is very 
sharply peaked in many dimensions. By doing a random walk rather than a
direct sampling, one stays where the integrand is large.
But this advantage is also a curse because it is not obvious
whether any given walk will converge to its equilibrium
distribution in the time available; this is the ergodic
problem. This aspect of simulation is experimental;
there are no useful theorems, only lots of controlled
tests, the lore of the practitioners, and occasional clean
comparisons with experimental data. Other subtleties of
these methods are how to pick the initial and boundary
conditions, determine error bars on the results, compute
long-range potentials quickly, and determine physical
properties \cite{Allen-Tildesley}.}.
Monte Carlo (Metropolis or Markov Chain) is a random
walk through phase space using rejections to achieve detailed
balance and thereby sample the density matrix. In particular, for the 
spatial integral we propose a displacement move of the position of a single 
timeslice $k$ of a single particle $i$ according to 
$\rr_{k,i}\to\rr_{k,i}+\ddelta(\eta-1/2)$ where $\eta$ is a uniform 
pseudo-random number in $[0, 1)$ and $\ddelta$ a fixed $d$-dimensional vector 
whose magnitude is chosen so to have acceptance ratios close to 1/2. 
Whereas, for the permutation sum, we propose a swap move of two randomly chosen 
particles through the {\sl L\'evy construction} of a Brownian bridge
(see Section V.G of Ref. \cite{Ceperley1995} and references therein). 
This particular sampling of the permutations sum will never be able
to change the winding number of 3 or more particles \cite{Ceperley1995},
so we will not be able to measure the superfluid fraction (see our code 
listing in the supplementary material).

Calling $\Phi=\lambda\phi_{\rm AQ}+\phi_{\rm HS}$ and realizing that this
total pair potential is everywhere continuous, we measure the following 
observables:
\begin{itemize}
\item[i.] The thermodynamic estimator \cite{Ceperley1995} for the potential 
energy is 
\bq
\calo=\calv=\frac{1}{M}\sum_{i=1}^M\sum_{k<l=1}^N\Phi(r_{kl,i}).
\eq 
\item[ii.] The thermodynamic estimator \cite{Ceperley1995} for the kinetic 
energy is
\bq
\calo=\calk=\frac{3N}{2\tau}-\frac{1}{M}\sum_{i=1}^M\sum_{k=1}^N
\frac{(\rr_{k,i}-\rr_{k,i-1})^2}{4\lambda\tau^2}.
\eq
So that the the total internal energy $\calo=\cale=\calk+\calv$.
\item[iii.] The virial estimator \cite{Ceperley1995} for the pressure
\bq
\calo=\calz=\frac{\beta}{3N}\left(2\calk-\frac{1}{M}\sum_{i=1}^M
\sum_{k<l=1}^Nr_{kl,i}\frac{d\Phi(r_{kl,i})}{dr}\right),
\eq 
where $\calz$ is the compressibility factor $\beta p/\rho$ with $p$ the pressure.
\end{itemize}
Note that in the classical HS fluid the potential energy vanishes identically. 
Not so in the quantum AQHS fluid where the repulsive spiked AQ effective pair 
potential will prevent any two HS from touching since the spike will produce 
a smooth repulsive potential profile in a neighborhood of the pair
classical contact $r_{ij}=\sigma^+$ that tends to keep apart the pair of particles: 
they will never be able to touch each other and will instead slip away one 
from the other contact. We expect this effect to affect the structure of the
fluid, its radial distribution function or its structure factor profiles.

Moreover, in the classical HS fluid the compressibility factor only depends
on the density or the packing fraction being {\sl athermal} 
\cite{Carnahan1969,Mansoori1971}. Not so in the quantum 
AQHS fluid where the presence of the effective affine potential makes the
product $\beta\Phi$ in the density matrix depend on temperature.     

So we already see how the two scenarios, the classical realm and the
(affine) quantum realm, present profound differences. In the next section we 
will present our computer experiment results.

\section{Numerical results}
\label{sec:nr}

We chose units such that: $\hbar=k_B=\sigma=1$. We will work with reduced 
quantities, like the reduced temperature $T^\star=k_BT/\lambda\sigma^{-2}$, 
and the reduced density $\rho^\star=\rho\sigma^3$.

In our simulations we had to keep under control 3 sources of error:
\begin{itemize}
\item[i.] approach to the {\sl continuum limit} $M\to\infty$ with $T$ constant, 
which, of course, can be reached only in a theoretical thinking. Here we will 
adopt the strategy to present all our results at fixed timestep $\tau=0.08$;
\item[ii.] approach to the {\sl thermodynamic limit} $N\to\infty$ with 
$\rho$ constant, which, again, can be reached only in a theoretical thinking.
This is known as the {\sl finite size error}. In order to make some progress
on a finite computer one usually employs periodic boundary conditions so that 
the simulation box and its periodic images fill up the whole space. Here we 
will present all our results at fixed $N=30$;
\item[iii.] the {\sl statistical error} which is inherently present in any 
Monte Carlo calculation. As the previous two sources of error also this one 
is unavoidable. We will here adopt the strategy of a faithful estimation
(through a careful determination of the autocorrelation time during the simulation
evolution) of the statistical error bar on any measure that we give. 
\end{itemize}

\begin{figure}[htbp]
\begin{center}
\includegraphics[width=8cm]{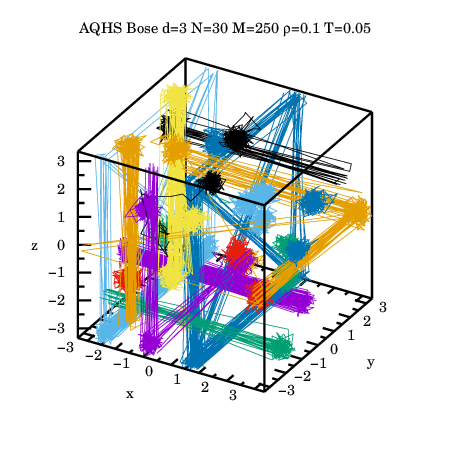}
\includegraphics[width=8cm]{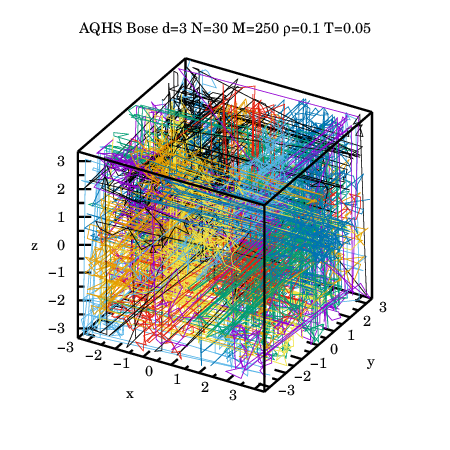}
\end{center}  
\caption{We show two snapshots of the paths of 30 AQHS made of 250 timeslices
each. At a density $\rho^\star=0.1$ and temperature $T^\star=0.05$. On the left, at the 
beginning of the simulation started with the paths at random non-overlapping 
positions repeated for all the $M=250$ beads of each of the $N=30$ paths.
On the right, at the end of the simulation when the random walk in the 
path integral reached equilibrium. During the simulation
we measured a kinetic energy per particle $\langle\calk\rangle/Nk_BT=37.0(3)$, 
a potential energy per particle $\langle\calv\rangle/Nk_BT=-6.72(2)$, and a
compressibility factor $\langle\calz\rangle=97.8(1)$.} 
\label{fig:snap}
\end{figure}

In Figure \ref{fig:snap} we show snapshots of the simulation box taken
at the beginning and at the end of the computer experiment.

In Table \ref{tab:nr} we summarize the numerical results obtained with our PIMC 
experiment in $d=3$ with 30 AQHS bosons of unit mass, keeping the timestep 
$\tau=0.08$ fixed. During the simulation we only measured thermodynamic 
quantities like the internal energy and the compressibility factor.

\begin{table}[hbt]
\caption{For a test case of $d=3$, $N=30$ boson AQHS particles with 
$m=1$ in a cubic periodic cell we measure, with PIMC at fixed $\tau=0.08$, 
at various values of the reduced temperature and of the reduced density, the 
total kinetic energy $\beta\langle\calk\rangle$, the total potential energy 
$\beta\langle\calv\rangle$, the total internal energy $\beta\langle\cale\rangle$,
and the compressibility factor $\langle\calz\rangle$.} 
\begin{center}
\begin{tabular}{||c||c|c||c|c|c|c||}
\hline\hline
$M$ & $k_BT$ & $\rho\sigma^3$ & $\langle\calk\rangle/Nk_BT$ & $-\langle\calv\rangle/
Nk_BT$ & $\langle\cale\rangle/Nk_BT$ & $\langle\calz\rangle$\\
\hline
\hline
250 & 0.05 & 0.10 & 37.0(3)  & 6.72(2)  & 30.3(3)   & 97.8(1) \\
250 & 0.05 & 0.05 & 17.0(3)  & 11.81(2) & 5.2(3)    & 42.9(2) \\
250 & 0.05 & 0.01 & 2.72(9)  & 6.690(6) & -3.96(9)  & 8.78(5) \\
\hline
31  & 0.40 & 0.10 & 4.23(5)  & 0.802(8) & 3.42(5)   & 12.06(1) \\
31  & 0.40 & 0.05 & 2.30(5)  & 1.464(6) & 0.83(5)   & 5.50(2) \\
31  & 0.40 & 0.01 & 1.67(3)  & 0.803(3) & 0.87(3)   & 2.02(1) \\
\hline
12  & 1.00 & 0.10 & 2.65(2)  & 0.345(3) & 2.30(2)   & 5.389(7) \\
12  & 1.00 & 0.05 & 1.87(2)  & 0.561(2) & 1.31(2)   & 2.875(6) \\
12  & 1.00 & 0.01 & 1.55(1)  & 0.293(2) & 1.26(1)   & 1.434(6) \\
\hline
8   & 1.50 & 0.10 & 2.15(2)  & 0.192(4) & 1.96(2)   & 3.9469(2) \\
8   & 1.50 & 0.05 & 1.73(1)  & 0.345(3) & 1.38(1)   & 2.296(4) \\
8   & 1.50 & 0.01 & 1.532(9) & 0.184(2) & 1.349(9)  & 1.305(4) \\
\hline\hline
\end{tabular}
\end{center}
\label{tab:nr}
\end{table}

\begin{figure}[htbp]
\begin{center}
\includegraphics[width=10cm]{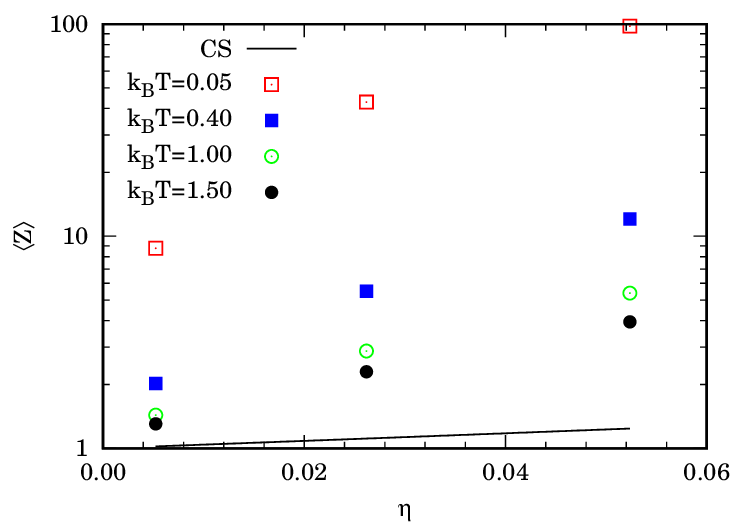}
\end{center}  
\caption{We show the compressibility factor $\langle\calz\rangle$ as a function of the 
packing fraction $\eta$ at various values of the temperature, from our numerical 
results of Table \ref{tab:nr}. For comparison, the Carnahan-Starling (CS) equation 
of state, $Z=(1+\eta+\eta^2-\eta^3)/(1-\eta)^3$, of classical HS \cite{Carnahan1969} 
is also shown as a continuous line. The logarithmic scale on the compressibility
factor axis is necessary because of the incompressibility nature of the AQHS fluid
at low temperature in his extremely quantum regime.} 
\label{fig:cf}
\end{figure}

In Figure \ref{fig:cf} we plot the results for the compressibility factor
presented in Table \ref{tab:nr} and compare them with the Carnahan-Starling 
(CS) equation of state (eos) \cite{Carnahan1969} for HS in the classical realm. 
As one can see from the figure at low temperature, when the quantum effects become 
important, our results indicate a rapid increase of the pressure at all densities.
And the monotonous increasing behavior of the pressure as a function of density
continues to hold at all temperatures. This increase reflects the tendency of
the fluid to crystallize at high densities. We also expect a supersolid
behavior in this cases, even if we are not yet ready to measure the 
superfluid fraction with our present treatment of the permutation sum in 
Eq. (\ref{eq:Omeasure}).
More importantly, we see that at high temperature when the path integral reduces
to a classical integral nonetheless our results still deviate largely from the
CS eos. This should be no surprise since the effective affine potential
$\phi_{\rm AQ}$ disappears only in the $\hbar\to 0$ mathematical limit, but not 
in the $\beta\to 0$ physical regime.

%

\section{Conclusions}
\label{sec:conclusions}

In classical hard-sphere fluids the particles can be variously decorated 
with adhesion, patchiness, active interaction sites, non-additivity, $\ldots$,
\cite{Fantoni04b,Fantoni05b,Fantoni06a,Fantoni06b,Fantoni06c,Fantoni07,Fantoni11e,
Fantoni13d,Fantoni14a,Fantoni17c,Fantoni18b,Fantoni23d} and all these aspects
play an important role in modellization in {\sl soft matter physics}.

Quantum hard-sphere fluids on the other hand lack, from the very beginning, of a 
precise definition. This is due to the fact that the momentum operator from 
the usual canonical-quantization of Dirac ceases to be self-adjoint in presence 
of forbidden regions or hard boundaries. We here proposed to use the {\sl dilation} 
operator of affine-quantization \cite{Fantoni23b} in order to reach a well defined
statistical mechanics model. 

In following this program we soon discovered some profound differences between
the hard-sphere fluid in the classical realm and the one in the quantum (affine) 
realm. In particular the property of the fluid to be athermal in its classical 
realm and and thermal in its quantum (affine) realm will survive even when taking 
the high temperature, classical, limit of the quantum (affine) 
fluid in its path integral formulation. In fact the effective affine potential
vanishes only in the $\hbar\to 0$ limit but not in the limit $\beta\to 0$.
In other words the two limits $\hbar\to 0$ and $\beta\to 0$ do not commute:
by taking first $\hbar\to 0$, at finite non-zero $\beta$, one delves into the 
classical domain leading to the Carnahan-Starling \cite{Carnahan1969} athermal 
equation of state for the HS; but taking first $\beta\to 0$, at finite $\hbar$,
one finds the classical limit of the AQHS fluid. These are two different things. 
A third further different approach to the HS quantum fluid would be to 
disregards the affine effective potential all together taking therefore 
$\Phi=\phi_{\rm HS}$ and employ this pair potential directly in the PIMC.
We suspect that most of the attempts made in literature to study the HS
quantum fluid used this third approach. But as we tried to explain this is 
not fully legitimate \cite{Giorgini2000,Sese2012,Serna2019}. 

We carried out some numerical experiments to measure the thermodynamic 
properties of the HS affine quantum fluid model and drew the equation of state 
in the pressure-density plane at various fixed values of the temperature. These 
extend to the quantum realm the ubiquitous Carnahan-Starling \cite{Carnahan1969} 
equation of state for hard-spheres valid only in the classical realm.

\section*{Author declarations}

\subsection*{Conflicts of interest}
None declared.

\subsection*{Data availability}
The data that support the findings of this study are available from the 
corresponding author upon reasonable request.

\subsection*{Funding}
None declared.

\bibliography{qhs-aq}

\end{document}